\documentclass[12pt]{article}
\usepackage{graphicx,multirow}
\usepackage{latexsym}
\usepackage[left=1in,right=1in,top=1.3in,bottom=1.3in]{geometry}
\usepackage{amsmath}
\usepackage{booktabs}
\usepackage{authblk}
\usepackage{color}
\usepackage{tikz}
\usepackage{graphicx}
\graphicspath{{figures/}}
\usepackage{subcaption}
\usepackage{amssymb,bm}
\usepackage{graphics}
\usepackage{amsthm}
\usepackage{float}
\usepackage{longtable}
\usepackage{amsmath}
\usepackage{algorithm} 
\usepackage{algpseudocode}
\usepackage{algorithmicx}
\usepackage{natbib}
\usepackage[colorinlistoftodos,textsize=tiny]{todonotes}
\usepackage{setspace}
\usepackage[nomarkers]{endfloat}

\DeclareMathOperator{\E}{\mbox{E}}
\DeclareMathOperator{\pr}{\mbox{pr}}

\usepackage{hyperref}
\usetikzlibrary{shapes}
\usetikzlibrary{arrows}
\definecolor{darkblue}{rgb}{0,0.4,0.9}
\definecolor{gray10}{rgb}{0.1,0.1,0.1}
\definecolor{gray20}{rgb}{0.2,0.2,0.2}
\definecolor{gray30}{rgb}{0.3,0.3,0.3}
\definecolor{gray40}{rgb}{0.4,0.4,0.4}
\definecolor{gray60}{rgb}{0.6,0.6,0.6}
\definecolor{gray80}{rgb}{0.8,0.8,0.8}
\definecolor{gray90}{rgb}{0.9,0.9,.9}
\definecolor{gray95}{rgb}{0.95,0.95,.95}
\definecolor{gray96}{rgb}{0.96,0.96,.96}
\definecolor{lgreen} {RGB}{180,210,100}
\definecolor{dblue}  {RGB}{20,66,129}
\definecolor{ddblue} {RGB}{11,36,69}
\definecolor{lred}   {RGB}{220,0,0}
\definecolor{nred}   {RGB}{224,0,0}
\definecolor{norange}{RGB}{230,120,20}
\definecolor{nyellow}{RGB}{255,221,0}
\definecolor{ngreen} {RGB}{98,158,31}
\definecolor{dgreen} {RGB}{78,138,21}
\definecolor{nblue}  {RGB}{28,130,185}
\definecolor{jblue}  {RGB}{20,50,100}
\definecolor{nnyellow}{RGB}{235,200,0}
\definecolor{purple}{RGB}{150, 0, 120}
\definecolor{sgGreen} {RGB}{20, 180, 50}
\definecolor{revised}{rgb}{0,0,0.9}

\newtheorem{theorem}{Theorem}

\usepackage{bbm}

\newcommand{\openr}{\hbox{${\rm I\kern-.2em R}$}}
\newcommand{\openn}{\hbox{${\rm I\kern-.2em N}$}}
\newcommand{\indep}{\rotatebox[origin=c]{90}{$\models$}}
\newcommand{\ind}{\mathbbm{1}}
\makeatletter
\newcommand*{\defeq}{\mathrel{\rlap{%
      \raisebox{0.3ex}{$\m@th\cdot$}}%
    \raisebox{-0.3ex}{$\m@th\cdot$}}%
  =}
\makeatother

\begin{document}

\title{Robust inference on the average treatment effect using the outcome highly adaptive lasso}

\author[1]{
  Cheng Ju$^{*}$
}

\author[2]{
 David Benkeser\thanks{Denotes an equal contribution to the manuscript}
}

\author[1]{
  Mark J. van der Laan
}

\date{}

\affil[1]{University of California, Berkeley, Berkeley, CA, USA}

\affil[2]{Emory University, Atlanta, GA, USA}

\maketitle

\begin{abstract}
Many estimators of the average effect of a treatment on an outcome require estimation of the propensity score, the outcome regression, or both. It is often beneficial to utilize flexible techniques such as semiparametric regression or machine learning to estimate these quantities. However, optimal estimation of these regressions does not necessarily lead to optimal estimation of the average treatment effect, particularly in settings with strong instrumental variables. A recent proposal addressed these issues via the outcome-adaptive lasso, a penalized regression technique for estimating the propensity score that seeks to minimize the impact of instrumental variables on treatment effect estimators. However, a notable limitation of this approach is that its application is restricted to parametric models. We propose a more flexible alternative that we call the outcome highly adaptive lasso. We discuss large sample theory for this estimator and propose closed form confidence intervals based on the proposed estimator. We show via simulation that our method offers benefits over several popular approaches. 

\end{abstract}


\newpage

\section{Introduction}

Across many fields, researchers are interested in the average effect of a treatment on an outcome. This ``treatment'' might correspond to a drug, a harmful exposure, or a policy intervention. Often, the treatment may not be randomized due to ethical or logistical reasons, which necessitates statistical methodology to address differences between those observed to take the treatment and those observed not to take the treatment \citep{rosenbaum1983central}. These differences are often accounted for through regression adjustment, either through estimation of the mean outcome given treatment and confounders (the \emph{outcome regression}), the probability of treatment given confounders (the \emph{propensity score}), or both. Many popular techniques for generating efficient estimates of the average treatment effect (ATE) utilize these regression estimates as an intermediate step. For example, the inverse probability of treatment weighted (IPTW) estimator, inverse weights the observed outcomes according to the inverse estimated probability of each observations receiving treatment. More involved procedures are available that use {\it both} the outcome regression and propensity score to generate an asymptotically efficient estimate of the average treatment effect. These efficient procedures include augmented inverse probability of treatment weighted (AIPTW) estimators and targeted minimum loss-based estimators (TMLEs). 

Commonly, the requisite regressions for estimation of the ATE are modeled using parametric techniques such as linear and logistic regression. However, misspecification of regression models can lead to extreme bias in estimates of the ATE \citep{kang2007demystifying}, which has led to growing interest in the use of nonparametric methods \citep{robins2006adaptive,hubbard2000nonparametric,hernan2010causal,farrell2015robust,kennedy2018nonparametric} and adaptive regression techniques to estimate the ATE \citep{van2006targeted,lee2010improving,karim2017estimating,wyss2018using,karim2018can}. In particular, the field of targeted learning has emerged as a paradigm for deriving formal statistical inference about estimated treatment effects when machine learning techniques are used to fit regressions \citep{van2011targeted,van2018targetedbook}. A particularly interesting machine learning algorithm in this context is the highly adaptive lasso (HAL, \citet{benkeser2016highly,van2017generally}). Under weak conditions, HAL achieves a fast convergence rate irrespective of the dimension of putative confounding variables. In fact, the convergence rate is sufficient to guarantee asymptotic efficiency of AIPTW estimators and TMLEs under weak conditions \citep{van2017generally}.

A key problem for any ATE estimation strategy is selecting confounders from a potentially larger set of measured variates. Recent studies have shown that including \emph{instrumental variables} -- variates that affect the propensity score, but not the outcome regression -- in the propensity score leads to inflation of the variance of the estimator of the average treatment effect relative to estimators that exclude such variables \citep{schisterman2009overadjustment,rotnitzky2010note,van2010collaborative,schneeweiss2009high,ju2017scalable}. \citet{shortreed2017outcome} proposed a data-driven approach to variable selection in an estimate of propensity scores. The proposal involves estimating the outcome regression using a generalized linear model and using the coefficients from this fit as evidence of how strongly each covariate is related to the outcome. The adaptive lasso \citep{zou2006adaptive} is used to fit the propensity score, where the penalty associated with each variable is proportional to the absolute value of the inverse of the variable's outcome regression coefficient. Thus, variables that are strongly related to the outcome (i.e., have a large coefficient in the outcome regression model) are penalized less, and are more likely to be included the propensity score. The resultant propensity score estimate is referred to as the {\it outcome adaptive lasso}. The authors use this estimated propensity score to construct an IPTW estimator of the ATE. 

In this work, we consider the implications of using the proposal of \citet{shortreed2017outcome}, but using HAL instead of parametric outcome regression and propensity score models. This seemingly straightforward extension turns out to have interesting implications for estimation and inference. In particular, the HAL representation of a function as a linear combination of tensor products of infinitesimal indicator basis functions leads to fine-grained screening for instruments. Specifically, the method screens for {\it instrumental basis functions}, the set of infinitesimal indicator basis functions that may be excluded from the outcome regression while still capturing its true form. The second interesting implication of utilizing Shortreed's method is that, by construction, the OHAL propensity score estimate may be inconsistent for the true propensity score when instrumental basis functions are present. Recent works have demonstrated that inference derived from procedures that na{\"i}vely use inconsistent nuisance estimators is often severely biased and further steps are needed to provide robust inference \citep{van2014targeted,benkeser2017doubly}. Building on these results, we propose a $n^{1/2}$-consistent estimator of the average treatment effect that has an asymptotic Normal sampling distribution and whose variance can be estimated in closed form irrespective of whether there are true instruments present in the data. We demonstrate the potential benefits of our approach via simulation. 

The remainder of the article is organized as follows. Section \ref{sec:back} provides background of this work, including identification of the ATE, estimation using TMLE, outcome-adaptive estimation of the propensity score, and estimation using HAL. Section \ref{sec:meth} proposes our estimator, describes its implementation, and discusses its weak convergence. Section \ref{sec:sim} provides an empirical study of the proposed estimator. We conclude with a discussion of possible future directions.

\section{Background}
\label{sec:back}

\subsection{Identification of average treatment effect}

Suppose we observe $n$ independent copies of the data unit $O \sim P_0$, which consists of $(W, A, Y)$, where $W$ is a $p$-dimensional vector of baseline covariates, $A$ is a binary treatment assignment, and $Y$ is an outcome of interest. Without loss of generality, we assume that $Y \in [0,1]$. We assume a nonparametric model $\mathcal{M}$ for $P_0$. Our interest is in evaluating the difference in average outcome if the entire population were assigned to receive $A = 1$ versus $A = 0$. Specifically, we follow \citet{pearl2009causality} and define a nonparametric structural equation model (NPSEM), $W=f_W(U_W)$, $A=f_A(W, U_A)$, $Y=f_Y(A, W, U_Y)$, where $f_W, f_A, f_Y$ are deterministic functions, and $U_W,U_A,U_Y$ are exogenous variables. The model assumes data are generated sequentially: the baseline covariates $W$ are generated based on $U_W$ and $f_W$; then, the treatment $A$ is generated $W$, $U_A$, and $f_A$; finally, the outcome $Y$ is generated based on $A$, $W$, $U_Y$, and $f_Y$. We consider intervening in this system to deterministically set $A$, rather than allowing $f_A$ to determine its value. This intervention generates $Y^{(a)} = f_Y(a, W_i, U_{Y,i}), a \in \{0, 1\}$, so-called counterfactual random variables. For $a = 0,1$, we denote by $\mathbb{P}_0^a$ the distribution of the counterfactual random variable $Y^{(a)}$. Our parameter of interest is $\E_{\mathbb{P}_0^1}[Y^{(1)}] - \E_{\mathbb{P}_0^0}[Y^{(0)}]$, which we refer to as the average treatment effect (ATE). 

The ATE is identifiable under the following assumptions: consistency, $Y_i = Y_i^{(A_i)}, \ i = 1,\dots,n$; no interference: $Y_i^{(A_i)}$ does not depend on $A_j$ for $i = 1,\dots,n$ and $j \ne i$; ignorability: $A_i \indep (Y_i^{(1)},Y_i^{(0)}) | W_i, \ i = 1, \dots, n$; positivity: $\mbox{pr}_{P_0}\{ 0 < \mbox{pr}_{P_0}(A = 1 \mid W) < 1\} = 1$. The first two assumptions are needed so that the counterfactual random variables are well defined. The ignorability condition states that there are no unmeasured confounders of $A$ and $Y$, while the positivity criterion states that every participant has a non-zero probability of receiving $A = 1$ and $A = 0$. If these assumptions hold, the average treatment effect is identified based on the observed data according to the G-computation identification \begin{equation} \label{gcomp}
    \E_{\mathbb{P}_0^1}\{Y^{(1)}\} - \E_{\mathbb{P}_0^0}\{Y^{(0)}\} = \E_{P_0}[\E_{P_0}\{Y \mid A = 1, W\} - \E_{P_0}\{Y \mid A = 0, W\}] \ .
\end{equation} 

\subsection{Locally efficient estimation of the average treatment effect using TMLE} \label{sec:local_eff}

For simplicity, in the remainder, we explicitly consider estimation of $\psi_0 = \E_{P_0}\{\E_{P_0}(Y \mid A = 1, W)\}$. A symmetric argument can be made for estimation of $\E_{\mathbb{P}_0^0}\{Y(0)\}$, and thus the ATE. Several nonparametric approaches have been proposed for estimating the ATE that use flexible estimates of the outcome regression and propensity score. Targeted minimum loss-based estimation (TMLE) is one such example. For each possible $w$, we denote by $\bar{Q}_n(w)$ an estimate of $\bar{Q}_0(w) := \E_{P_0}(Y \mid A = 1, W = w)$, the true outcome regression at $w$. Similarly, we denote by $\bar{G}_n$ an estimate of the propensity score $\bar{G}_0(w) := \pr_{P_0}(A = 1 \mid W = w)$. Using these estimates, a TMLE is computed by defining a particular parametric working model. Because $Y \in [0,1]$, we may use a logistic working model for the outcome regression, $\{\bar{Q}_{n,\epsilon} = \mbox{expit}[\mbox{logit}(\bar{Q}_n) + \epsilon H_n] : \epsilon \in \mathbb{R}\}$, where $H_n(W_i) := 1/\bar{G}_n(W_i).$ Note that this defines a logistic regression model with offset given by the logit of the initial estimator and single covariate $H_n(W)$. We also note that the use of this particular submodel is not restricted to binary outcomes; in particular, any real-valued outcome can be scaled to the unit interval by subtracting the minimum value of $Y$ and dividing by the range of $Y$, so that this same model can be used \citep{gruber2010targeted}. A maximum likelihood estimator $\epsilon_n$ of $\epsilon$ is found using data from observations with $A_i = 1$ (e.g., using iteratively re-weighted least-squares) and the TMLE is $\psi_n^* := \frac{1}{n} \sum_{i=1}^n \bar{Q}_{n, \epsilon_n}(W_i).$ 

Study of the stochastic properties of TMLEs is facilitated through a linearization of the G-computation parameter, which involves the efficient influence function, a key object in efficiency theory \citep{bickel1998efficient}. Given an outcome regression $\bar{Q}$, a propensity score $\bar{G}$, and a marginal cumulative distribution function of $W$, $Q$, we define the nonparametric efficient influence function \begin{align}
D(O_i \mid \bar{Q}, \bar{G}, Q) &:= \frac{A_i}{\bar{G}(W_i)} \{Y_i - \bar{Q}(W_i)\} + \bar{Q}(W_i) - \int \bar{Q}(w) dQ(w) \ . \label{EIF_def}  
\end{align}
We remark that the efficient influence function is a key ingredient in many causal effect estimators. Indeed, the choice of submodel used in the TMLE procedure described above is directly related to the form of the efficient influence function. The efficient influence function is also important for estimating equations-based estimators and one-step estimators, which are asymptotically equivalent with TMLE (for more discussion, see \citet{van2018targeted}).

We denote by $Q_n$ the empirical cumulative distribution function of $W_1,\dots,W_n$. If the outcome regression and propensity score estimators are such that $D(\cdot \mid \bar{Q}_n, \bar{G}_n, Q_n)$ falls in a $P_0$-Donsker class with probability tending to one and $\int \{D(o \mid \bar{Q}_n, \bar{G}_n, Q_n) - D(o \mid \bar{Q}_0, \bar{G}_0, Q_0)\}^2 dP_0(o)$ converges in probability to zero then \begin{equation} \begin{aligned} \label{expansion}
&\psi_n^* - \psi_0 = \frac{1}{n} \sum_{i = 1}^n D(O_i \mid \bar{Q}_0, \bar{G}_0, Q_0) + R_{n} + o_{\text{p}}(n^{-1/2}) \ , 
\end{aligned}
\end{equation}
where the second-order remainder is \begin{align}
R_{n} &:= \int \left[ \{\bar{Q}_{n,\epsilon_n}(w) - \bar{Q}_0(w)\} \left\{ \frac{\bar{G}_n(w) - \bar{G}_0(w)}{\bar{G}_n(w)} \right\}  \right] dQ_0(w) \ .\label{remainder_def}
\end{align}
This term can be shown to be asymptotically negligible, in the sense that $R_{n} = o_{\text{p}}(n^{-1/2})$, provided \begin{align}
\int \{\bar{Q}_n(w) - \bar{Q}_0(w)\}^2 dQ_0(w) &= o_{\text{p}}(n^{-1/4}) \ , \  \mbox{and} \notag \\ 
\int \{\bar{G}_n(w) - \bar{G}_0(w)\}^2 dQ_0(w) &= o_{\text{p}}(n^{-1/4}) \ . \label{prop_quarterrate}
\end{align}

If indeed $R_{n} = o_{\text{p}}(n^{-1/2})$, then $\psi_{n}^* - \psi_0 = \frac{1}{n}\sum_{i=1}^n D(O_i \mid \bar{Q}_0, \bar{G}_0, Q_0) + o_{\text{p}}(n^{-1/2})$. Thus, the central limit theorem implies that the centered and scaled estimator, $n^{1/2}(\psi_{n}^* - \psi_0)$ converges in distribution to a mean-zero Gaussian random variable with variance $\sigma^2_0 = \mbox{var}_{P_0}\{D(O \mid \bar{Q}_0, \bar{G}_0, Q_0)\} = \int D(o \mid \bar{Q}_0, \bar{G}_0, Q_0)^2 dP_0(o)$.
The asymptotic variance $\sigma^2_0$ may be consistently estimated by \begin{equation} \label{var_estimator} \begin{aligned}
	\sigma_n^2 &= \mbox{var}_{P_n}\{D(O \mid \bar{Q}_n, \bar{G}_n, Q_n)\} = \int D(o \mid \bar{Q}_n, \bar{G}_n, Q_n)^2 dP_n(o) \\
	&= \frac{1}{n} \sum_{i=1}^n D(O_i \mid \bar{Q}_n, \bar{G}_n, Q_n)^2 \ .
	\end{aligned}
\end{equation} 
The Wald-style confidence interval \begin{equation} \label{standard_ci}
	\biggl(\psi_{n}^* - z_{1-\alpha/2} \frac{\sigma_n}{n^{1/2}} \ , \ \psi_{n}^* + z_{1-\alpha/2} \frac{\sigma_n}{n^{1/2}}\biggr)
\end{equation}
will have asymptotic coverage probability no smaller than $1 - \alpha$.

\subsection{Outcome-adaptive estimation of propensity scores}

It is common in many applications that relevant baseline covariates must be selected from a larger set of variables. Therefore, an important consideration in any analysis is deciding what variables should comprise $W$. The ignorability assumption suggests that, at a minimum, $W$ should include all variables that are related both with the treatment $A$ and the outcome $Y$. However, there is also potential benefit in including variables related only to the outcome \citep{zhang2008improving,moore2009covariate}, and there is risk in including variables related only to the treatment \citep{schisterman2009overadjustment}. Moreover, in finite samples it is often beneficial to remove true confounding variables that cause extreme values in the propensity score \citep{petersen2012diagnosing}. This has motivated many proposals for automated variable selection for estimation of the ATE \citep{rotnitzky2010note,schneeweiss2009high,van2010collaborative,ju2017scalable,ju2018collaborative}. 

An interesting approach to automated variable selection was discussed by \citet{shortreed2017outcome}. Their proposal can be summarized as follows. First, we define a main terms parametric regression model for the outcome regression, e.g., $\{ \mbox{logit}[\bar{Q}_{\alpha,\eta}(a,w)] = \alpha_0 + \eta a + w^\top \alpha : (\alpha_0, \eta, \alpha) \in \mathbb{R}^{p+2}\}$ and estimate $(\alpha_0, \eta, \alpha)$ via maximum likelihood. We denote by $\hat{\alpha}_j$ the coefficient for variable $W_j$ in the outcome regression, and define $\hat{\alpha} := (\hat{\alpha}_j : j = 1,\dots,p)$. Next, we define a logistic regression model for the propensity score, e.g., $\{\mbox{logit}\{\bar{G}_{\beta}(1,w)\} = \beta_0 + w^\top \beta : (\beta_0, \beta) \in \mathbb{R}^{p+1}\}$ and estimate $(\beta_0, \beta)$ via the adaptive lasso \citep{zou2006adaptive} with penalty weight for $\beta_j$ given by $|\hat{\alpha}_j|^{-\gamma}$, where $\gamma$ is a user-supplied tuning parameter. The authors discuss using the resultant propensity score estimate to construct a stabilized IPTW estimate of the average treatment effect, though in principle, the outcome regression and IPTW estimates could be used in a locally efficient procedure like TMLE. 

\subsection{Highly adaptive lasso}

The highly adaptive lasso (HAL) estimator is a semiparametric regression estimator that is consistent for the true regression function assuming only that the true regression function has finite variation norm \citep{benkeser2016highly,van2017generally}. Moreover, a TMLE constructed based on HAL estimates of the outcome regression and propensity score will generally satisfy the requisite regularity conditions discussed in section \ref{sec:local_eff}. HAL is therefore an appealing choice for use when estimating the average treatment effect in settings where flexible regression models are needed. 

A HAL regression of an outcome $Z$ on covariates $X$ is achieved by mapping $X$ into a set of indicator basis functions. If $X$ is univariate, we generate a basis expansion for a typical observation $x$ as $\phi^*_n(x) = (\phi^*_1(x), \dots, \phi^*_n(x))^{\top}$, where for $j = 1,\dots,n$, $\phi_j^*(x) = I(x \ge X_j)$. An $L_1$-penalized regression of the outcome is fit using these basis functions, with the optimal $L_1$-norm chosen via cross-validation. In a single dimension, HAL is equivalent with zero-order trend filtering \citep{kim2009ell}. In higher dimensions, tensor-product basis functions are included in the basis expansion, e.g., if $X = (X_1,X_2)$, we generate first-order basis functions as well as second-order basis functions of the form $\phi_j^*(x) = I(x_1 \ge X_{1j}, x_2 \ge X_{2j})$ for $j \in \{1, \dots, n\}$. We use $\bm{\phi}_n(x)$ to denote the $p$-vector of basis functions evaluated at a typical observation $x$. The intuition for the particular choice of basis functions used by HAL is as follows. For each $\lambda$ on the lasso solution path, the corresponding HAL estimate has been shown to be a minimum loss-estimator (MLE) over the class of functions with variation norm no larger than the $L^1$-norm of the coefficient vector. For each $\lambda$, this is a Donsker class, which implies a fast convergence rate of the MLE. Oracle properties of cross-validation imply that selecting the variation norm using cross-validation does not adversely affect the convergence rate. For more details pertaining to HAL, we refer readers to the original publications. 

Given the formulation of HAL as a regression of an outcome onto a set of (binary indicator) variables, HAL provides a natural extension of the proposal of \citet{shortreed2017outcome} for outcome-adaptive propensity score estimation. In the next section, we discuss implications for estimation and inference. 

\section{Methodology}
\label{sec:meth}

\subsection{Outcome highly adaptive lasso propensity estimation}

The reliance of \citet{shortreed2017outcome} on parametric regression models limits its applicability settings where more flexible models are required. However, because the highly adaptive lasso in a way is just a special case of the usual lasso, it is straightforward to accommodate outcome-adaptive weights in the HAL-based estimation of the propensity score. An algorithm for generating a propensity score estimate is as follows. First, we compute the HAL estimate of the outcome regression, using $Y$ as outcome and $W$ as predictors in the subgroup of participants with $A = 1$. We denote by $\hat{\alpha}$ the vector of HAL-estimated coefficients associated with the HAL basis functions. Next, we fit HAL using the treatment indicator $A$ as outcome, and $W$ as features, but using outcome-adaptive weights. Specifically, the HAL model for the propensity score is $
\{\mbox{logit}\{\bar{G}_{\beta_0, \beta}(w)\} := \beta_0 + \beta^{\top} \bm{\phi}_n(w) : \beta \in \mathbb{R}^{d_n + 1}\},$ where $d_n$ is the dimension of $\bm{\phi}_n$, which is no larger than $n(2^p - 1)$. Fitting HAL corresponds to the optimization problem of finding $(\hat{\beta}_0(\lambda), \hat{\beta}(\lambda)) := \mbox{argmin}_{\beta_0,\beta} \sum_{i=1}^n \mbox{log}[ \bar{G}_{\beta_0, \beta}(W_i)^{A_i} \{1 - \bar{G}_{\beta_0, \beta}(W_i) \}^{1-A_i} ] + \lambda \sum_{j=1}^p |\hat{\alpha}_j|^{-\gamma} | \beta_{j}|,$ where $\gamma$ is a user-supplied tuning parameter. The penalty parameter $\lambda$ is selected via cross-validation. We call the resultant fit the outcome highly adaptive lasso (OHAL). 

\begin{figure}
\centering
\includegraphics[width = \textwidth]{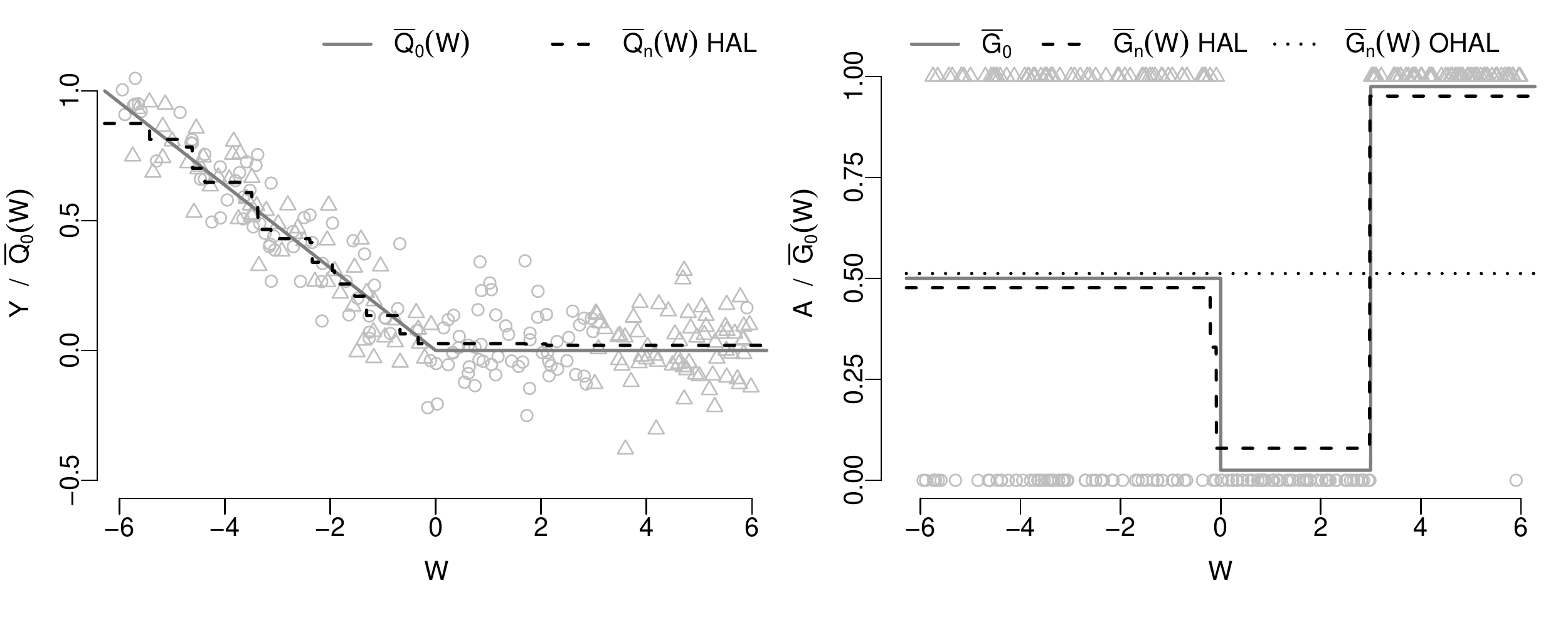}
\caption{Illustration of HAL vs. OHAL propensity score estimation on a data set of $n = 250$. $W$ from a Uniform(-6, 6) distribution. The true outcome regression and propensity score are shown in gray. HAL fits are shown as dashed lines, and the OHAL fit is shown as a dotted line.}
\label{simple_dgp}
\end{figure}

An interesting observation on OHAL propensity estimation is that individual basis functions, rather than entire variables are penalized in the fit.  Thus, we could say that OHAL screens for {\it instrumental basis functions}, those basis functions that parameterize directions in the covariate space along which the outcome regression does not vary. Therefore, the approach provides an extremely granular means of screening for instruments and removing them to avoid possibly extreme propensity score estimates. To illustrate the potential benefit, consider the data set illustrated in Figure \ref{simple_dgp}. 

In this example, $W$ is not an instrumental variable and, due to the overall linear trend in the outcome regression, would likely not be labeled as such by common screening procedures. However, when $W$ is viewed as a collection of infinitesimal indicator basis functions, any basis function of the form $\phi_w(u) = I(u \ge w)$ for $w \ge 0$ is an instrumental variable because the outcome regression is constant over this region. Moreover, there are extreme propensity score values for these same values of $W$, which may cause erratic behavior of causal effect estimators. Thus, to improve performance of causal effect estimators, we may wish to remove these basis functions from the propensity score fit. This is precisely the goal of our proposed approach. Theoretically, there may be relevant benefits to this approach in both small and large samples. In small samples, avoiding extreme propensity score estimates may yield more stable estimates of treatment effects. Considering large samples we note that OHAL will converge pointwise to $\bar{G}(\bar{Q}_0) = \bar{G}_0(w)$ for $w < 0$ and $\pr_{P_0}(A = 1 \mid W \ge 0)$ for $w \ge 0$. Indeed, we see in the single data analysis, OHAL fits a constant function of $W$ that is close to $\bar{G}_0(w)$ for $w < 0$ and in between the extreme values of $\bar{G}_0(w)$ for $w > 0$. We recall that the form of the asymptotic variance of a nonparametric efficient estimator is $\mbox{var}_{P_0}\{D(O \mid \bar{Q}_0, \bar{G}_0, Q_0)\}$ and that the efficient influence function $D$ involves dividing by $\bar{G}_0$. Because $\bar{G}_0(w)$ is near to 0 for some $w$ the efficient variance may be quite large. On the other hand, because $\bar{G}(\bar{Q}_0)$ pools the propensity score over all $w > 0$, we might expect that a locally efficient estimator based on OHAL rather than, e.g., HAL will lead to improvements in asymptotic variance relative to a locally efficient estimator.

\subsection{Inference when using OHAL}

Unfortunately, there are generally repercussions with respect to inference when considering locally efficient estimators that are based on inconsistent estimates of either the outcome regression or propensity score \citep{van2014targeted,benkeser2017doubly}. In particular, the remainder (\ref{remainder_def}) may not be asymptotically negligible if instrumental variables (or instrumental basis functions) are present in $\bar{Q}_0$, due to the fact that OHAL is estimating $\bar{G}(\bar{Q}_0)$ rather than $\bar{G}_0$. Thus, the remainder term may contribute to the first-order behavior of a locally efficient estimator. If $\bar{Q}_n$ is a well-specified MLE in a finite-dimensional regression model for $\bar{Q}_0$, then locally efficient estimators will still have an asymptotic Normal distribution, though the variance estimator (\ref{var_estimator}) and confidence interval (\ref{standard_ci}) will generally no longer be valid. In the case that $\bar{Q}_n$ is based on an adaptive algorithm, such as HAL, the repercussions on inference are severe: asymptotically, the confidence interval (\ref{standard_ci}) has zero coverage probability and has been shown to behave poorly in finite samples \citep{benkeser2017doubly}. \citet{van2014targeted} proposed TMLE methods that yield confidence intervals that are robust to inconsistent estimation of one of the outcome regression or propensity score. An alternative TMLE approach was proposed and evaluated in \citet{benkeser2017doubly}. 

\subsection{Proposed estimator}

We propose a TMLE based on a HAL estimate of the outcome regression and an OHAL estimate of the propensity score. Because HAL is used, our outcome regression estimator is consistent under weak conditions. However, the OHAL propensity score estimator is consistent for $\bar{G}(\bar{Q}_0)$, which may or may not equal the true propensity score. This all depends on whether instrumental basis functions are present. Because the analyst is unlikely to have a-priori knowledge of whether or not such basis functions are present, we make use of the results of \citet{benkeser2017doubly} to develop an estimator and confidence intervals that have sound theoretical behavior in either situation without needing to know which situation is true. Our proposed estimator is efficient when no instrumental basis functions are present and may be super efficient otherwise. In both cases, the proposed estimator has a closed-form influence function, which we can use to construct robust Wald-style confidence intervals. Relative to a standard TMLE, this estimator requires estimation of two additional, univariate regression functions, and includes a two- rather than one-step targeting procedure. 

The proposed TMLE estimator can be implemented as follows. \\
{\it Obtain initial estimates of nuisance parameters}
\begin{enumerate}
\item[(i)] use HAL to obtain estimate $\bar{Q}_n$ of $\bar{Q}_0$;
\item[(ii)] use OHAL based on $\bar{Q}_n$ to obtain estimate $\bar{G}_n$ of $\bar{G}(\bar{Q}_0)$;
\item[(iii)] use HAL to obtain estimate $\bar{G}_{r,1,n}$ of the ``reduced-dimension'' regression $\bar{G}_{r,1,0} := \pr_{P_0}\{A = 1 \mid \bar{Q}_0\}$ by fitting HAL with outcome $A_i$ and univariate covariate $\bar{Q}_n(W_i)$, for $i = 1,\dots,n$;
\item[(iii)] use HAL to obtain estimate $\bar{G}_{r,2,n}$ of the ``reduced-dimension'' regression $\bar{G}_{r,2,0} := \E_{P_0}[\{A - \bar{G}(\bar{Q}_0)(W)\}/\bar{G}(\bar{Q}_0)(W) \mid \bar{Q}_0]$ by fitting HAL with outcome $\{A_i - \bar{G}_n(W_i)\}/\bar{G}_n(W_i)$, for $i = 1,\dots, n$.
\end{enumerate}
{\it Iterative targeting of outcome regression} \begin{enumerate}
\item[(iv)] define $H_{r,n}(W_i) = \frac{\bar{G}_{r,2,n}(W_i)}{\bar{G}_{r,1,n}(W_i)}$ and $H_n(W_i) = \frac{1}{\bar{G}_n(W_i)}$, $i = 1, \dots, n$; 
\item[(v)] fit a logistic regression using outcome $Y_i$, offset $\mbox{logit}\{\bar{Q}_n(W_i)\}$ and single covariate $H_{r,n}(W_i)$ using observations with $A_i = 1$ ; call the estimate of the univariate parameter of this regression $\epsilon_{r,n}$ and set $\mbox{logit}\{\bar{Q}_{1,n}(W_i)\} = \mbox{logit}\{\bar{Q}_n(W_i)\} + \epsilon_{r,n} H_{r,n}(W_i)$;
\item[(vi)] fit a logistic regression using outcome $Y_i$, offset $\mbox{logit}\{\bar{Q}_{1,n}(W_i)\}$ and single covariate $H_{r,n}$ using observations with $A_i = 1$; call the estimate of the univariate parameter of this regression model $\epsilon_{2,n}$, and define the targeted outcome regression as $\bar{Q}_n^*(W_i) = \mbox{expit}[\mbox{logit}\{\bar{Q}_{1,n}(W_i)\} + \epsilon_{2,n} H_n(W_i)]$.
\end{enumerate}
Repeat steps (iv)-(vi) above until $\frac{1}{n}\sum_{i=1}^n D(O_i \mid \bar{Q}_n^*, \bar{G}_n, Q_n) < c_n$ and $\frac{1}{n}\sum_{i=1}^n D_r(O_i \mid \bar{Q}_n^*, \bar{G}_{r,n}) < c_n$, where $c_n = o_{\text{p}}(n^{-1/2})$. In our simulation, we used $c_n = 1/\{n^{1/2}\mbox{log}(n)\}$. \\
{\it Compute plug-in estimator} \begin{enumerate}
\item[(vii)] the TMLE is $\psi_n^* := \frac{1}{n} \sum_{i=1}^n \bar{Q}_n^*(W_i)$.
\end{enumerate}
In the supplement, we describe an alternative implementation that avoids iteration. We found no practical difference between the estimators in our simulation. 

The motivation for the ``reduced-dimension'' regressions $\bar{G}_{r,0} := (\bar{G}_{r,1,0}, \bar{G}_{r,2,0})$ can be explained as follows. If $\bar{G}_0(\bar{Q}_0) \ne \bar{G}_0$, i.e., instrumental basis functions are present, then the remainder $R_n$ can be viewed as a nonparametric plug-in estimator of 0, which we may express as $R(\bar{Q}_n)$, where $R$ maps from the outcome regression model space to $\mathbb{R}$. In general, plug-in estimators based on nonparametric regression functions, like HAL, are not $n^{1/2}$-consistent estimators of their target (c.f., certain sieve estimators \citep{chen2007large}). For example, the plug-in estimator $\frac{1}{n} \sum_{i=1}^n \bar{Q}_n(W_i)$ of $\psi_0$, based on HAL is not generally $n^{1/2}$-consistent. This fact motivates the use of TMLE, where we estimate an additional regression, namely the propensity score, and we modify our initial estimate of $\bar{Q}_0$ based on the estimated propensity score. A parallel can be drawn to the situation above: $R(\bar{Q}_n)$ is not a $n^{1/2}$-consistent estimator of zero and so we use a step of TMLE that involves estimation of additional regression quantities $\bar{G}_{r,0}$ that we use to modify our initial estimate of $\bar{Q}_0$. The result is that $R(\bar{Q}_n^*)$ is a $n^{1/2}$-consistent estimator of zero. 

\subsection{Statistical inference}
For a typical observation $O_i$, and given outcome regression $\bar{Q}$ and ``reduced-dimension'' regressions $\bar{G}_{r} := (G_{1,r}, G_{2,r})$, we define \begin{align*}
D_{r}(O_i \mid \bar{Q}, \bar{G}_r) &:= A_i \frac{\bar{G}_{r,1}(W_i)}{\bar{G}_{r,2}(W_i)} \{ Y_i - \bar{Q}(W_i) \} \ . 
\end{align*}
Note that $\bar{G}_r$ implicitly is indexed by a given propensity score, though we omit this notation for simplicity. In the theorem, we assume the index is set to the OHAL limit $\bar{G}(\bar{Q}_0)$. We now present a theorem describing the weak convergence of $\psi_n^*$. 

\begin{theorem}
Under regularity conditions explicitly stated in the appendix, \[
	\psi_{n}^* - \psi_0 = \frac{1}{n}\sum_{i=1}^n \left\{D(O_i \mid \bar{Q}_0, \bar{G}(\bar{Q}_0), Q_0) - D_{r}(O_i \mid \bar{Q}_0, \bar{G}_{r,0})\right\} + o_{\text{p}}(n^{-1/2}) \ ,
\]
and $n^{1/2}(\psi_n^* - \psi_0)$ converges weakly to a mean-zero Normal variate with variance \[
	\tau^2_0 := \emph{var}_{P_0}\{D(O \mid \bar{Q}_0, \bar{G}(\bar{Q}_0), Q_0) - D_r(O \mid \bar{Q}_0, \bar{G}_{r,0}) \} \ . 
\]
\end{theorem}

Note that when the OHAL limit $\bar{G}_0(\bar{Q}_0)$ equals the true propensity score $\bar{G}_{r,2,0}(w) = 0$ for all $w$ and thus $D_r(o \mid \bar{Q}, \bar{G}_{r,0} = 0)$ for all $o$. In words, when there are no instrumental variables nor instrumental basis functions, $\psi_n^*$ is a nonparametric efficient estimator of $\psi_0$. On the other hand, if there are such basis functions, then $\psi_n^*$ is still a $n^{1/2}$-consistent estimator of $\psi_0$, though there is a first-order contribution to its influence function stemming from the intentional inconsistent estimation of $\bar{G}_0$. 

In our numeric studies, we evaluate Wald-style confidence intervals based on two different standard error estimates. The first is a typical influence-function-based estimate, $\tau_{n}^2 := \mbox{var}_{P_n}\{D(O \mid \bar{Q}_n, \bar{G}_n, Q_n) - D_r(O \mid \bar{Q}_n, \bar{G}_{n,r}) \}$. The second is a \emph{partially} cross-validated influence-function-based estimator. The HAL outcome regression and OHAL propensity score regression each utilize $V$-fold cross-validation to select the lasso penalty parameter. Thus, after these regressions are fit, we have $V$ lasso fits available for each of the outcome regression and outcome-adaptive propensity score, one for each training fold. For $i = 1,\dots, n$, we denote by $\bar{Q}_{n,v}$ the HAL-estimated outcome regression estimated in the $v$-th training fold. Given an observation $O_i$ in the $v$-th validation fold, we say that $\bar{Q}_{n,v}(W_i)$ is a \emph{partially} cross-validated estimate of $\bar{Q}_0(W_i)$. We say \emph{partially} cross-validated since observation $i$ was used to select the lasso tuning parameter. Similarly, we denote by $\bar{G}_{n,v}$ the outcome-adaptive HAL propensity score estimate obtained in the $v$-th training fold. Again, this estimator is \emph{partially} cross-validated, since observation $i$ was used both to construct the HAL outcome regression estimator and to select the tuning parameter for the OHAL fit. Finally, we denote by $\bar{G}_{n,r,v}$ the estimate of the reduced-dimension regression obtained in the $v$-th training fold (i.e., based on $\bar{Q}_{n,v}$ and $\bar{G}_{n,v}$). For $v = 1, \dots, V$, we denote by $P_{n,v}$ the empirical distribution of observations in the $v$-th validation fold and define \begin{equation} \label{cv_var_estimate}
	\tau_{n,\text{CV}} := \frac{1}{V} \sum_{v=1}^V \mbox{var}_{P_{n,v}}\{ D(O \mid \bar{Q}_{n,v}, \bar{G}_{n,v}, Q_n) - D_r(O \mid \bar{Q}_{n,v}, \bar{G}_{n,r,v}) \} \ . 
\end{equation}
In spite of this estimator's reliance on cross-validation, it is obtained with essentially no additional computational burden beyond that needed to obtain $\psi_n^*$, since it merely recycles cross-validated lasso fits needed to compute $\psi_n^*$ itself. 

\section{Simulation}
\label{sec:sim}

We designed a simulation to highlight the two potential benefits of our proposed methodology. First, our proposal uses flexible estimators of the outcome regression and can adapt to complex underlying true regression functions. In these settings, we expect benefit of our approach relative to approaches that utilize misspecified parametric regression models. On the other hand, more flexible regression methodologies can be used to capture these relationships. However, the second benefit of our proposal is that we use OHAL to estimate the propensity score, which may be expected to offer benefits in settings with instrumental variables and/or instrumental basis functions. So relative to other nonparametric approaches, we also expect to see improvements. Therefore, we designed a simulation study that included both non-linear relationships and instrumental variables and compared our proposal to several other approaches in this setting. 

We simulated one thousand data sets at each of three sample sizes, $n \in \{100, 500, 1000\}$. We drew $W = (W_1, W_2, W_3, W_4)$ by independently sampling each component from a Uniform(-1, 1), Bernoulli(0.5), Uniform(-1, 1), and Uniform(0,1) distribution, respectively. The treatment was drawn from a Bernoulli distribution with $\bar{G}_0(1,W) = \mbox{logit}^{-1}\{0.5  - W_3 + 2 W_3 W_2 - 2.5 W_4\}$. The outcome was drawn from a Bernoulli distribution with $\bar{Q}_0(A,W) = \mbox{logit}^{-1}\{-2 W_1 \ind(W_1 > -1/2) - W_3 + 2 W_2 W_3 + A\}$. The average treatment effect implied by this data generating distribution is approximately 0.20. 

We studied several estimators of the average treatment effect. First, we considered two G-computation (also known as standardization estimators \citep{robins1986new}) estimators based on parametric models. The first was based on a parsimonious, correctly-specified logistic regression model for the outcome regression. The second was based on a misspecified, main terms logistic regression model. We also considered two inverse probability of treatment weighted (IPTW) estimators, with one again based on a parsimonious, correctly-specified logistic regression and one based on a misspecified, main terms model for the propensity score. For both the G-computation and IPTW estimator, we view the estimators based on correctly specified regression models as idealized benchmarks, since in practice it is often difficult to correctly specify a parametric regression model. Confidence intervals for these estimators were constructed using a percentile-based nonparametric bootstrap based on 500 bootstrap samples. 

We also implemented Shortreed's estimator of the average treatment effect, where the outcome regression was inconsistently estimated via a main terms logistic regression and the propensity score was estimated using the outcome adaptive lasso. Treatment effect estimators based on this approach should remove deleterious effects of the instrumental variable ($W_4$), but will yet suffer from incorrect specification of the outcome regression. We implemented the IPTW estimator proposed in Shortreed et al, in addition to a TMLE based on these regression fits. Confidence intervals for these estimators were constructed using the bootstrap approach recommended in Shortreed \citep{efron2014estimation}. 

For a nonparametric treatment effect estimator, we used TMLE based on HAL-estimated outcome regression and HAL-estimated propensity score. This estimator should properly account for non-linearity, but will not remove the instrumental variable, leaving the possibility of extreme estimated propensity scores. On the other hand, our proposed TMLE that uses OHAL estimated propensity score is expected to simultaneously account for non-linearity, avoid extreme estimated propensity scores, and maintain low bias in spite of the inconsistent propensity score estimate. Two types of confidence intervals were computed for both TMLE HAL and TMLE OHAL. The first was based on the estimated influence function, as in (\ref{var_estimator}); the second was based on the cross-validated estimate of the influence function, as in (\ref{cv_var_estimate}).

We compared these estimators on their Monte Carlo-estimated bias, standard error, and mean squared-error. We also compared the coverage of nominal 95\% confidence intervals, as well as the median width of the confidence interval. 

We found that, as expected, the benchmark estimators were essentially unbiased and had low mean squared-error, with the correctly-specified G-computation estimator performing slightly better than the IPTW estimator (top two rows, Table \ref{sim_results_table1}). On the other hand, the estimators based on misspecified parametric models exhibited large bias, resulting in large MSE. The benefits of the outcome-adaptive lasso are seen in the improved standard error of the estimators based on an OAL propensity estimate relative to those that do not penalize the inclusion of instrumental variables. However, the bias of both OAL-based estimators are large owing to the misspecification of the relevant regressions. TMLE based on HAL had relatively large bias in small sample sizes, but bias improved in larger samples. By comparison, our proposed TMLE based on OHAL had low bias in all sample sizes. In the smallest sample size, our proposed estimator had improved bias and standard error relative to TMLE based on HAL. In larger sample sizes, our estimator had improved bias. Overall, TMLE with OHAL had MSE 66\%, 75\%, and 75\% that of the TMLE based on HAL for sample sizes $100, 500,$ and $1000$, respectively.

\begin{table}[ht]
\centering
\begin{tabular}{lccc|ccc|ccc}
 & \multicolumn{3}{c}{Bias} & \multicolumn{3}{c}{Standard error} & \multicolumn{3}{c}{Mean squared-error} \\ 
 \hfill $n = $ & 100 & 500 & 1000 & 100 & 500 & 1000 & 100 & 500 & 1000 \\
  \cmidrule{2-4}\cmidrule{5-7}\cmidrule{8-10}
  GCOMP GLM (C) & -0.03 & 0.09 & 0.02 & 1.10 & 1.08 & 1.03 & 1.21 & 1.17 & 1.06 \\ 
  IPTW GLM (C) & -0.02 & 0.05 & -0.05 & 1.38 & 1.32 & 1.25 & 1.90 & 1.75 & 1.57 \\ 
  GCOMP GLM (M) & 0.56 & 1.34 & 1.82 & 1.09 & 1.10 & 1.03 & 1.51 & 3.02 & 4.36 \\ 
  IPTW GLM (M) & 0.61 & 1.43 & 1.90 & 1.22 & 1.17 & 1.11 & 1.85 & 3.41 & 4.85 \\ 
  IPTW OAL (M) & 0.59 & 1.42 & 1.90 & 1.15 & 1.15 & 1.10 & 1.68 & 3.33 & 4.82 \\ 
  TMLE OAL (M) & 0.66 & 1.48 & 1.98 & 1.17 & 1.15 & 1.09 & 1.81 & 3.51 & 5.11 \\ 
  TMLE HAL & 0.58 & 0.91 & 0.73 & 1.27 & 1.27 & 1.19 & 1.95 & 2.44 & 1.93 \\ 
  DRTMLE OHAL & 0.09 & 0.39 & 0.10 & 1.13 & 1.30 & 1.21 & 1.29 & 1.83 & 1.47 \\ 
   \hline
\end{tabular}
\caption{Monte Carlo bias (scaled by $n^{1/2}$), standard error (scaled by $n^{1/2}$), and mean squared-error of the estimators (scaled by $n$) at three sample sizes. Abbreviations: GCOMP = G-computation, IPTW = inverse probability of treatment-weighted estimator, GLM = generalized linear model, (C) = estimator based on correctly-specified regression, (M) = estimator based on misspecified regression, OAL = outcome-adaptive lasso, TMLE = targeted minimum loss-based estimation, HAL = highly adaptive lasso, DRTMLE = doubly robust TMLE, OHAL = outcome highly adaptive lasso.}
\label{sim_results_table1}
\end{table}

The benchmark estimators had near nominal coverage at all sample sizes (Table \ref{sim_results_table2}). Estimators based on misspecified parametric models had poor coverage. Surprisingly, the estimators based on OAL had near nominal coverage in large sample sizes. Because of the bias of these estimators, we suspect that coverage is due to conservative standard error estimates. Indeed, we find that in large samples, these confidence intervals tended to be considerably wider than the other intervals considered. For both nonparametric estimators, we found that the cross-validated standard error estimates provided better coverage in small samples than their non-cross-validated counterparts. Confidence intervals for our proposed estimator based on cross-validated standard error estimates had near nominal coverage at each sample size, while those based on the TMLE that used HAL provided coverage less than the nominal level. 

\begin{table}[ht]
\centering
\begin{tabular}{lccc}
 & \multicolumn{3}{c}{CI coverage (median width)} \\ 
 \hfill $n = $ & 100 & 500 & 1000 \\
 \hline
  GCOMP GLM (C) & 94.3 (0.42) & 93.6 (0.18) & 95.6 (0.13) \\ 
  IPTW GLM (C) & 93.6 (0.52) & 93.5 (0.23) & 95.8 (0.16) \\ 
  GCOMP GLM (M) & 89.5 (0.42) & 73.9 (0.19) & 60.1 (0.13) \\ 
  IPTW GLM (M) & 90.9 (0.48) & 75.6 (0.2) & 61.4 (0.14) \\ 
  IPTW OAL (M) & 89.4 (0.45) & 91.2 (0.28) & 95.2 (0.24) \\ 
  TMLE OAL (M) & 86.1 (0.45) & 89.8 (0.28) & 94.2 (0.24) \\ 
  TMLE HAL & 75.1 (0.34) & 74.4 (0.16) & 81.5 (0.12) \\ 
  TMLE HAL (CV-SE) & 93.1 (0.48) & 86.2 (0.21) & 91.9 (0.15) \\ 
  DRTMLE OHAL & 88.1 (0.38) & 86.4 (0.18) & 93.2 (0.14) \\ 
  DRTMLE OHAL (CV-SE) & 96.7 (0.48) & 91.9 (0.21) & 96.4 (0.15) \\  
   \hline
\end{tabular}
\caption{Monte Carlo coverage probability of nominal 95\% confidence intervals and median width of confidence intervals. Abbreviations: CI = confidence interval; CV-SE = cross-validated standard error; for others, see Table \ref{sim_results_table1}}
\label{sim_results_table2}
\end{table}

\section{Discussion}
\label{sec:discuss}

OHAL provides a new tool for flexibly estimating the propensity score while simultaneously accounting for fine-grained variable selection. Our simulation results point to potential benefits in utilizing this approach over existing nonparametric efficient estimators in some situations. In future work, we may wish to investigate the benefit of using OHAL for estimating the causal effect of a treatment administered over several time points, where extreme propensity scores are commonly encountered in practice. Another avenue for future research is to explore methods for adaptively tuning the regularization for HAL outcome regression. The outcome regression in OHAL plays a particularly important role as it contributes to the estimation of the average treatment effect both through its role as an estimator of the true outcome regression, as well as through its role in the propensity score estimation. Thus, errors in tuning the outcome regression may easily propagate down to the estimate of the average treatment effect. We have begun experimentation with recursive tuning procedures, but so far have found no change in the estimators' performance. Another important consideration is selection of the hyper-parameter $\gamma$ in OHAL. In our simulations, we simply set $\gamma = 1$, but it is possible that with more careful tuning of $\gamma$, the performance of estimators based on OHAL may be further improved. 

\section{Acknowledgement}

This project was supported by NIH grant R01 AI074345-08 and UC Berkeley Superfund Research Program P42 ES004705-29.

\bibliographystyle{plainnat}
\bibliography{references}

\clearpage

\section*{Appendix}

\subsection{Proof of Theorem 1}
In the proof, we adopt the shorthand notation $P_0 f$ to denote $\int f(o) dP_0(o)$ for any $P_0$-integrable function $f$. Consequently, using $P_n$ to denote the empirical distribution function of $n$ observations of $O$, we have $P_n f = n^{-1} \sum_{i=1}^n f(O_i)$. Our proof relies on the following regularity conditions: \begin{enumerate}
	\item[A1.] $D(\cdot \mid \bar{Q}_n, \bar{G}_n, Q_n)$ falls in a $P_0$-Donsker class with probability tending to one and $P_0\{D(\cdot \mid \bar{Q}_n, \bar{G}_n, Q_n) - D(\cdot \mid \bar{Q}_0, \bar{G}_0(\bar{Q}_0), Q_0)\}^2$ converges in probability to 0. 
	\item[A2.] $\int \{\bar{G}_n(w) - \bar{G}_0(\bar{Q}_0)(w)\}^2 dQ_0(w) = o_{\text{p}}(n^{-1/4})$.
	\item[A3.] $\int \{\bar{Q}_n(w) - \bar{Q}_0(w)\}^2 dQ_0(w) = o_{\text{p}}(n^{-1/4})$.
	\item[A4.] $\int \{\bar{G}_{r,2,0n}(w) / \bar{G}_{r,1,0n}(w) - \bar{G}_{r,2,0}(w) / \bar{G}_{r,1,0}(w)\}^2 dQ_0(w)= o_{\text{p}}(n^{-1/4})$
	\item[A5.] $\int \{\bar{G}_{r,2,n}(w) / \bar{G}_{r,1,n}(w) - \bar{G}_{r,2,0}(w) / \bar{G}_{r,1,0}(w)\}^2 dQ_0(w)= o_{\text{p}}(n^{-1/4})$
	\item[A6.] $D_r(\cdot \mid \bar{Q}_n, \bar{G}_{r,n})$ falls in a $P_0$-Donsker class with probability tending to one and $P_0\{D_r(\cdot \mid \bar{Q}_n, \bar{G}_{r,n}) - D_r(\cdot \mid \bar{Q}_0, \bar{G}_{r,0})\}^2$ converges in probability to 0. 
\end{enumerate}

We note that by design our estimators $\bar{Q}_n^*$, $\bar{G}_n$, $\bar{G}_{r,n}$ satisfy \begin{align}
&P_n D(\cdot \mid \bar{Q}_n^*, \bar{G}_n, Q_n) = o_{\text{p}}(n^{-1/2}) \label{eif_eqn1} \ , \mbox{and} \\
&P_n D_r(\cdot \mid \bar{Q}_n^*, \bar{G}_{r,n}) = o_{\text{p}}(n^{-1/2}) \label{eif_eqn2} \ . 
\end{align}
We note also that the efficient influence function is doubly robust in the sense that $P_0 D(\cdot \mid \bar{Q}_0, \bar{G}, Q_0) = 0$, for any choice of $\bar{G}$. 

We begin with an exact second-order linearization of the parameter, $\psi_n^* - \psi_0 = -P_0 D(\cdot \mid \bar{Q}_n^*, \bar{G}_n, Q_n) + R_n$. We add and subtract $(P_n - P_0) D(\cdot \mid \bar{Q}_0, \bar{G}_0(\bar{Q}_0), Q_0)$ and rearrange terms. Noting the double-robustness of the efficient influence function and (\ref{eif_eqn1}), we arrive at the equation \[
	\psi_n^* - \psi_0 = P_n D(\cdot \mid \bar{Q}_0, \bar{G}_0(\bar{Q}_0), Q_0) + R_n + M_n \ , 
\]
where $M_n = (P_n - P_0)\{D(\cdot \mid \bar{Q}_n, \bar{G}_n, Q_n) - D(\cdot \mid \bar{Q}_0, \bar{G}_0(\bar{Q}_0), Q_0)\}$. By assumption A1, $M_n = o_{\text{p}}(n^{-1/2})$. Thus, it remains to examine the behavior of the remainder. Now, \begin{align*}
R_n &= P_0 \left\{ \frac{\bar{G}_n - \bar{G}_0}{\bar{G}_n} (\bar{Q}_n - \bar{Q}_0)  \right\} = P_0 \left\{ \frac{\bar{G}_n - \bar{G}_0}{\bar{G}_0(\bar{Q}_0)} (\bar{Q}_n - \bar{Q}_0)  \right\} + R_{1,n} \\
&= P_0 \left\{ \frac{\bar{G}_0(\bar{Q}_0) - \bar{G}_0}{\bar{G}_0(\bar{Q}_0)} (\bar{Q}_n - \bar{Q}_0)  \right\} + R_{1,n} + R_{2,n}
\end{align*}
where $R_{1,n} = P_0 [ (\bar{Q}_n - \bar{Q}_0) \{\bar{G}_n - \bar{G}_0(\bar{Q}_0)\} \{\bar{G}_0(\bar{Q}_0) - \bar{G}_n\}/\{\bar{G}_n \bar{G}_0(\bar{Q}_0)\} ]$ and $R_{2,n} = P_0 [ (\bar{Q}_n - \bar{Q}_0)  \{\bar{G}_n - \bar{G}_0(\bar{Q}_0)\} / \bar{G}_0(\bar{Q}_0)  ]$. By assumption A2 and A3, $R_{1,n} = o_{\text{p}}(n^{-1/2})$ and $R_{2,n} = o_{\text{p}}(n^{-1/2})$. Now, allowing some abuse of our shorthand notation, \begin{align*}
P_0 \left\{ \frac{\bar{G}_0(\bar{Q}_0) - \bar{G}_0}{\bar{G}_0(\bar{Q}_0)} (\bar{Q}_n - \bar{Q}_0)  \right\} &= -P_0 \left\{ \frac{A - \bar{G}_0(\bar{Q}_0)}{\bar{G}_0(\bar{Q}_0)} (\bar{Q}_n - \bar{Q}_0)  \right\} \\
&= -P_0 \left\{ G_{r,2,0n} (\bar{Q}_n - \bar{Q}_0)  \right\} \\
&= P_0 \left\{ \frac{A}{\bar{G}_{r,1,0n}} \bar{G}_{r,2,0n} (\bar{Q}_n - \bar{Q}_0)  \right\} \ , 
\end{align*}
where $\bar{G}_{r,2,0n}(w) := \E_{P_0}[\{A - \bar{G}_0(\bar{Q}_0)(W)\} / \bar{G}_0(\bar{Q}_0) \mid \bar{Q}_n(W) = \bar{Q}_n(w), \bar{Q}_0(W) = \bar{Q}_0(w)]$ and $\bar{G}_{r,1,0n}(w) := \E_{P_0}\{A \mid \bar{Q}_n(W) = \bar{Q}_n(w), \bar{Q}_0(W) = \bar{Q}_0(w)\}$. We note that \[
P_0 \left\{ \frac{A}{\bar{G}_{r,1,0n}} \bar{G}_{r,2,0n} (\bar{Q}_n - \bar{Q}_0)  \right\} = -P_0 D_r(\cdot \mid \bar{Q}_n, \bar{G}_{r,n}) + R_{3,n} + R_{4,n} \ , 
\]
where $R_{3,n} = P_0 \{ A (\bar{G}_{r,2,0n} / \bar{G}_{r,1,0n} - \bar{G}_{r,2,0} / \bar{G}_{r,1,0}) (\bar{Q}_0 - \bar{Q}_n)  \}$ and $R_{4,n} = P_0\{ A (\bar{G}_{r,2,0} / \bar{G}_{r,1,0} - \bar{G}_{r,2,n} / \bar{G}_{r,1,n}) (\bar{Q}_0 - \bar{Q}_n) \}$. Assumptions A3-A5 imply that $R_{3,n} + R_{4,n}= o_{\text{p}}(n^{-1/4})$. We can now add and subtract $(P_n - P_0) D_r(\cdot \mid \bar{Q}_0, \bar{G}_{r,0})$ and rearrange terms. Noting the double-robustness of $D_r$ and (\ref{eif_eqn2}), we have \[
-P_0 D_r(\cdot \mid \bar{Q}_n, \bar{G}_{r,n}) = -P_n D_r(\cdot \mid \bar{Q}_n, \bar{G}_{r,n}) + M_{1,n} \ , 
\]
where $M_{1,n} = (P_n - P_0)\{ D_r(\cdot \mid \bar{Q}_n, \bar{G}_{r,n}) - D_r(\cdot \mid \bar{Q}_0, \bar{G}_{r,0}) \}$. Assumption A6 implies that $M_{1,n} = o_{\text{p}}(n^{-1/2})$, which completes the proof.  

\end{document}